\documentclass[11pt]{article}
\usepackage[T1]{fontenc}
\usepackage[utf8]{inputenc}
\usepackage[margin=1in]{geometry}
\usepackage{amsmath,amssymb}
\usepackage{graphicx}
\usepackage{booktabs}
\usepackage{xcolor}
\usepackage{caption}
\usepackage[hidelinks]{hyperref}

\definecolor{wfpurple}{HTML}{3D3580}

\title{\textbf{weightflow}: declarative, recipe-aware survey weighting in R}
\author{Juan Pablo Ferreira\\[2pt]
\small Instituto Nacional de Estad\'istica (INE), Uruguay\\
\small Facultad de Ciencias Econ\'omicas y de Administraci\'on,\\
\small Universidad de la Rep\'ublica, Uruguay\\
\small \texttt{juanpablo.ferreira@fcea.edu.uy}}
\date{\today}

\newcommand{\code}[1]{\texttt{#1}}
\newcommand{\pkg}[1]{\textsf{#1}}

\begin{document}
\maketitle

\begin{abstract}
\noindent
Producing analysis weights for a complex survey requires a sequence of
hierarchical adjustments (resolving unknown eligibility, discarding
out-of-scope units, restoring within-household selection probabilities,
correcting for nonresponse, and calibrating to known population totals), after
which design-consistent variances must account for the fact that several of
those adjustments were themselves estimated from the sample. Existing R tools
cover parts of this workflow, but none expresses the \emph{whole} cascade as a
single, auditable object, nor propagates the variability of every stage into the
replicate weights. We present \pkg{weightflow}, a dependency-free (base R)
package that builds survey weights through a declarative, pipeable,
\pkg{tidymodels}-style API: a recipe is defined lazily as a chain of
\code{step\_*()} adjustments and then estimated with \code{prep()}. Separating
\emph{definition} from \emph{application} makes the process reproducible and
auditable, and lets a rescaling bootstrap and a delete-a-PSU jackknife re-apply
the entire recipe on each replicate, so the replicate weights carry the
uncertainty of every estimated stage of the cascade, not only of the final
calibration. The package implements
raking, post-stratification and linear/GREG calibration (with truncated and
exponential distances, ridge penalisation, and domain-partitioned and
integrative variants), model-assisted (Wu--Sitter) calibration, weighting-class
and machine-learning response-propensity adjustments with cross-fitting, and
representativity (R-)indicators as an automatic diagnostic. Weights and
recipe-aware replicate weights bridge to the \pkg{survey} and \pkg{srvyr}
packages for design-based inference. We describe the methodology and the design
of the API, validate the calibration and variance results against \pkg{survey},
and illustrate the full cascade both on a bundled multistage sample and on real
household-survey microdata (the Uruguayan ECH), where it recovers a known poverty
rate with design-based uncertainty.
\end{abstract}


\section{Introduction}\label{sec:intro}

\begin{figure}[t]
\centering
\includegraphics[width=0.85\textwidth]{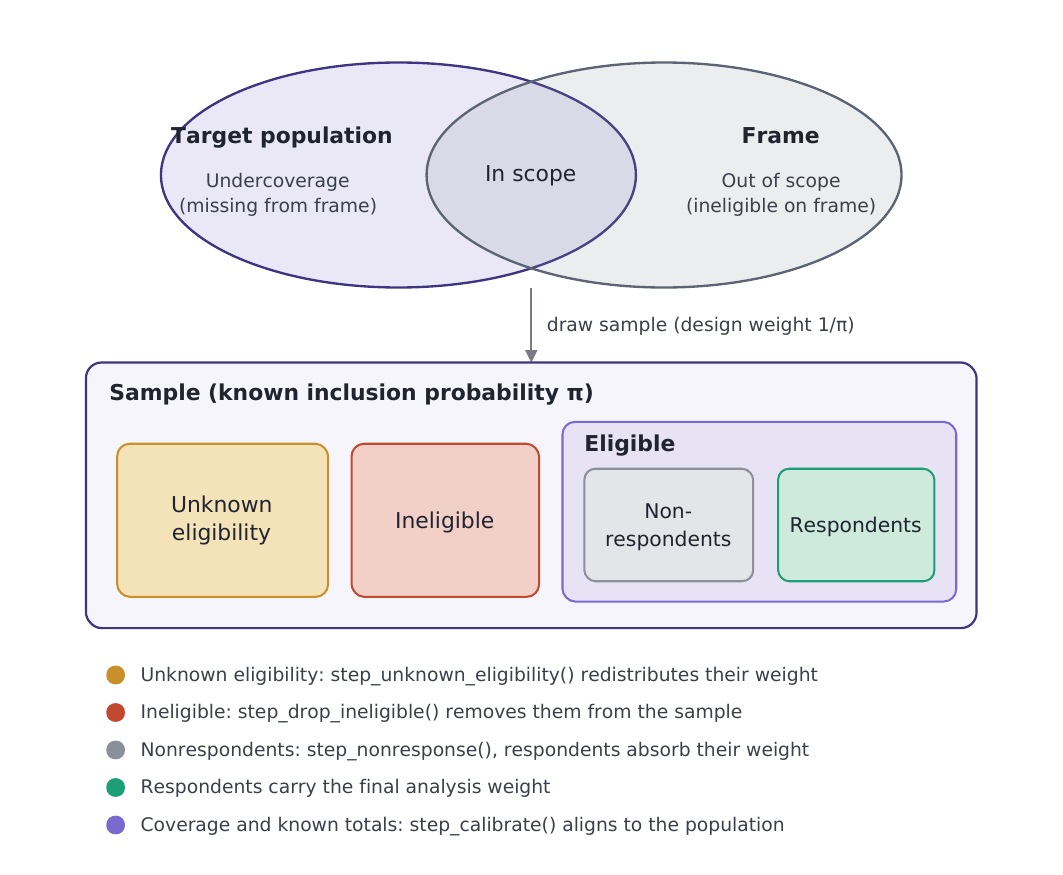}
\caption{The weighting problem as a map. The target population and the sampling
frame overlap (coverage error); the sample is drawn from the frame and then
classified by eligibility and, among the eligible, by response. Each region is
corrected by a specific \code{step\_*()} of the cascade.}
\label{fig:map}
\end{figure}

Weights turn a probability sample into statements about a population. Let
$U=\{1,\dots,N\}$ denote the target population, the universe of $N$ units we want
to describe, and let $y_i$ be the value of a study variable for unit $i$; the
goal is to estimate finite-population quantities such as the total
$Y=\sum_{i\in U} y_i$. A probability sample $s\subseteq U$ is drawn so that each
unit $i$ enters $s$ with a known inclusion probability $\pi_i$ and carries a
design (base) weight $w_i^{0}=1/\pi_i$. The Horvitz--Thompson estimator of $Y$ is
then $\hat Y_{HT}=\sum_{i\in s} w_i^{0}\,y_i$, unbiased under ideal conditions
\cite{sarndal1992}. In practice those conditions fail: frames have coverage
errors, some sampled units are found ineligible or of unknown eligibility, only
one person may be selected within a household, and nonresponse reduces the
effective sample. The design weights are therefore refined through a
\emph{cascade} of adjustments, each estimated from the data, and only then
calibrated to auxiliary population totals. Figure~\ref{fig:map} places each of
these departures on the map from the target population to the responding sample,
and names the adjustment that corrects it.

The R ecosystem offers mature tools for \emph{parts} of this process. The
\pkg{survey} package \cite{lumley2004} is the reference for estimation and for
calibration on fixed auxiliary totals; \pkg{sampling} draws samples and
calibrates design weights; \pkg{icarus} mirrors the SAS \textsc{Calmar} macro
for calibration; \pkg{surveysd} and \pkg{svrep} build calibrated and replicate
weights. What is missing is a tool that treats the \emph{entire} weighting
pipeline (eligibility, selection, nonresponse and calibration) as one explicit
object that can be read top to bottom, audited, and re-executed, and whose
variance estimator reflects every estimated stage rather than only the final
calibration.

In practice this cascade is usually implemented as a sequence of scripts (design
weights in one file, eligibility adjustments in another, calibration in a third,
variance estimation somewhere else) communicating through intermediate files.
The methodology is real but \emph{implicit}: scattered across file order, column
names and undocumented tweaks, so that two analysts can implement the same
strategy differently and audits, updates and methodological reviews become
unnecessarily hard; when the responsible methodologist leaves, the knowledge
often leaves too. \pkg{weightflow} makes the entire strategy a single explicit
object: it does not merely compute weights, it \emph{documents how they are
constructed}. The final weights are only one output of that object; the recipe
itself, a reproducible record of every methodological decision, is the
institutional asset.

\pkg{weightflow} fills that gap. Its contributions are:
\begin{enumerate}
\item A \textbf{declarative recipe API} that separates the lazy
  \emph{definition} of the weighting pipeline from its \emph{application},
  making the whole cascade reproducible and auditable
  (Section~\ref{sec:api}).
\item \textbf{Recipe-aware variance}: a rescaling bootstrap
  \cite{raowu1988} and a delete-a-PSU jackknife that re-run the complete recipe
  on each replicate, so nonresponse and calibration uncertainty enter the
  replicate weights, which bridge to \pkg{survey}/\pkg{srvyr}
  (Section~\ref{sec:variance}).
\item A \textbf{comprehensive, dependency-free} implementation of calibration
  (raking, post-stratification, linear/GREG, with truncated and exponential
  distances, ridge penalisation, domain-partitioned and integrative variants),
  model-assisted calibration \cite{wusitter2001}, and weighting-class and
  machine-learning nonresponse adjustments with cross-fitting
  (Sections~\ref{sec:calib}--\ref{sec:nr}).
\item Automatic diagnostics, including Kish's design effect and
  \textbf{representativity (R-)indicators} \cite{schouten2009}
  (Section~\ref{sec:diag}).
\end{enumerate}
The package targets national statistical offices (NSOs) and survey
practitioners. It builds on established adjustments rather than proposing new
point estimators, but its recipe-aware variance propagates the uncertainty of
\emph{every} estimated stage of the cascade, which no existing package does, and
it exposes the whole process through a single, transparent and reproducible
interface.

\section{The weighting cascade}\label{sec:cascade}

\begin{figure}[t]
\centering
\includegraphics[width=0.42\textwidth]{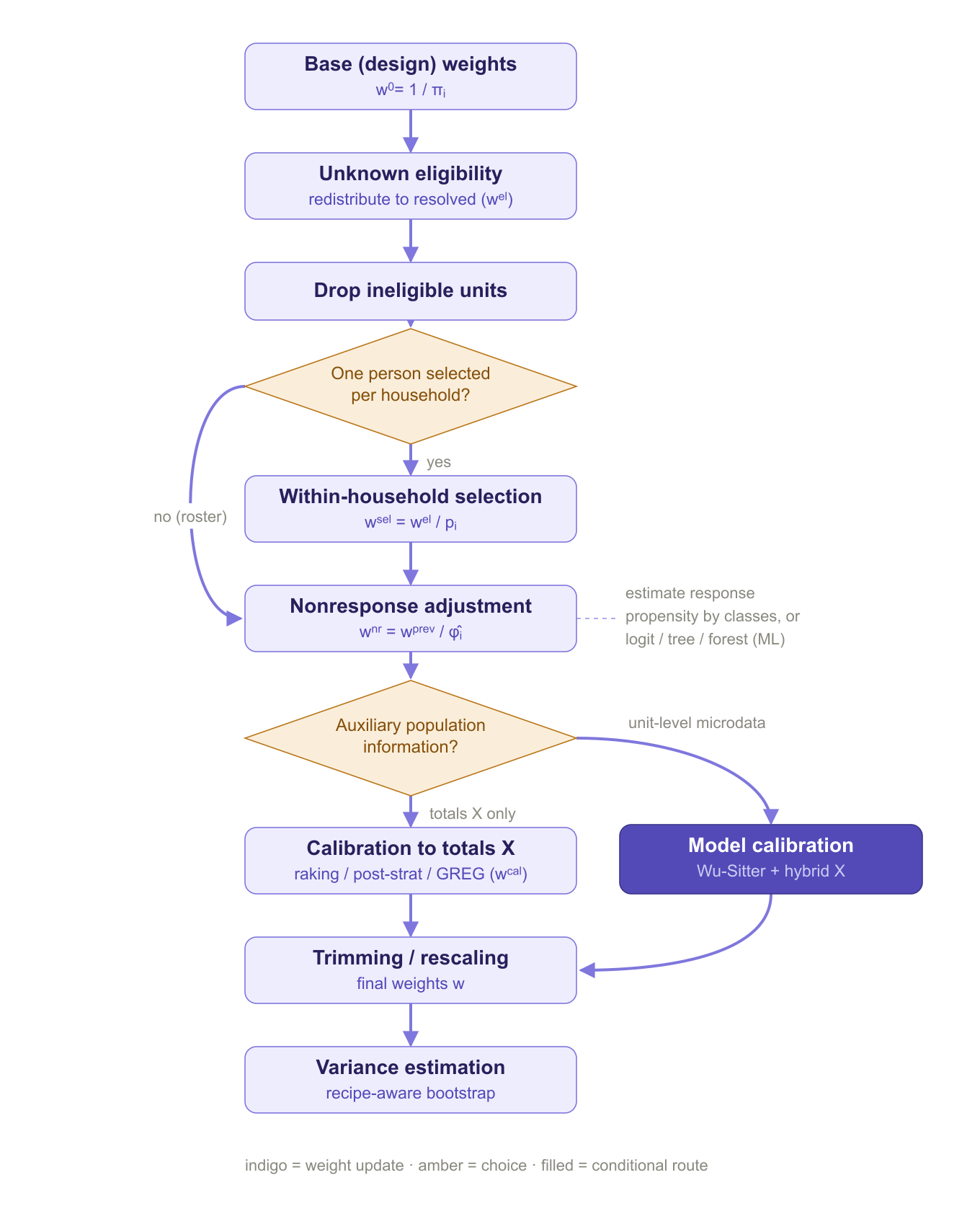}
\caption{The staged weighting process, from the design weight to the final
analysis weight. Each stage corrects a specific departure from the ideal design.}
\label{fig:flow}
\end{figure}

Starting from the design weight, each stage addresses one departure from the
ideal design (Figure~\ref{fig:flow}). The order matters, and every stage
modifies the current weight by a multiplicative factor.

\begin{figure}[t]
\centering
\includegraphics[width=0.72\textwidth]{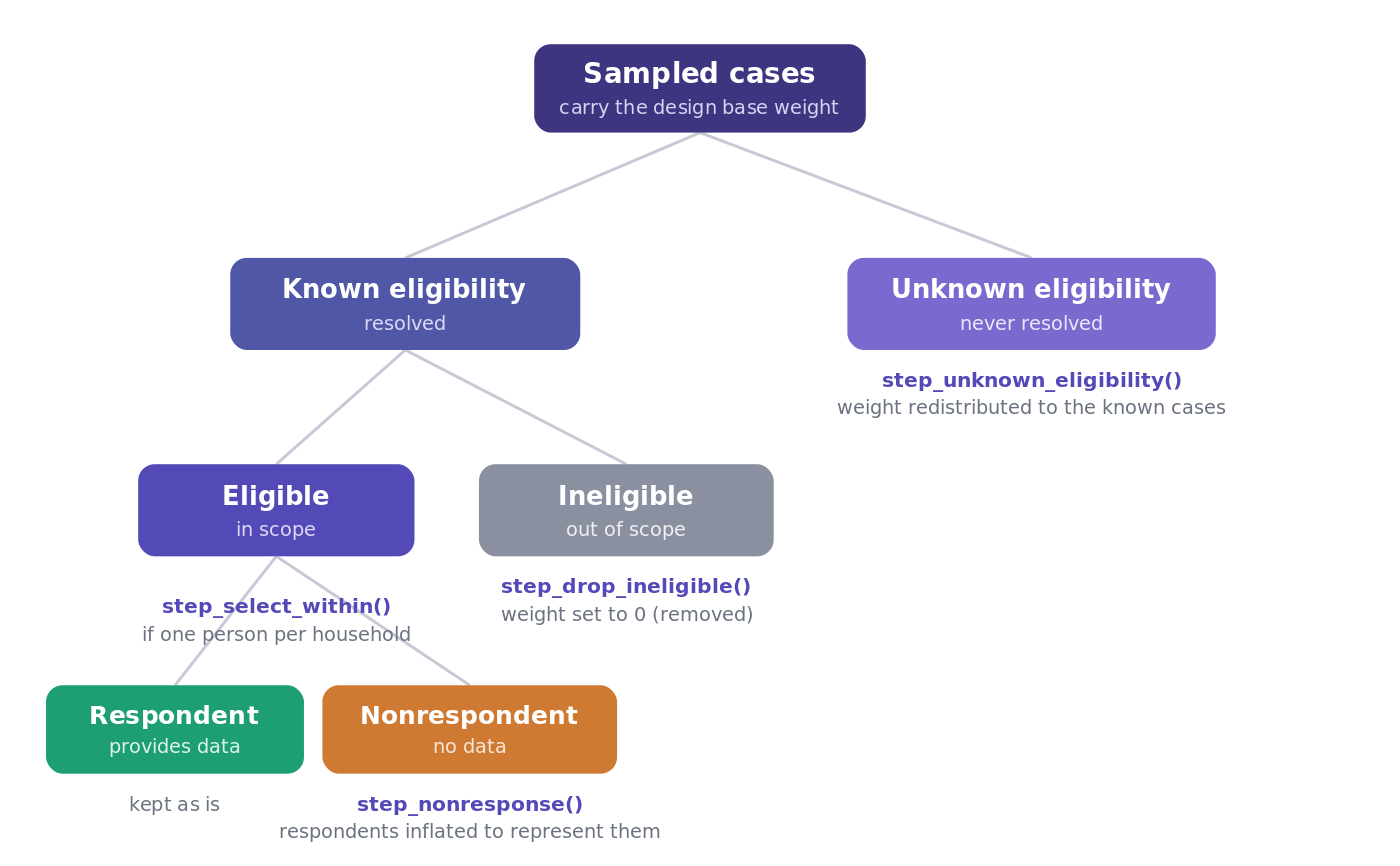}
\caption{The disposition tree. Each branch maps to a step of the cascade;
unknown-eligibility weight is redistributed, ineligibles are dropped, and
respondents are inflated to represent nonrespondents.}
\label{fig:tree}
\end{figure}

\paragraph{Disposition.}
Before any factor is computed, each sampled case is classified: by eligibility
(eligible / ineligible / unknown) and, among the eligible, by response
(respondent / nonrespondent), following the survey-methodology standard
\cite{valliant2018} and the AAPOR final-disposition categories
\cite{aapor2016}. Each disposition is handled by a different adjustment
(Figure~\ref{fig:tree}).

\paragraph{Unknown eligibility.}
Cases that are never resolved cannot be classified as in- or out-of-scope.
Ignoring them assumes they represent nobody; treating them all as eligible
overstates the population. \pkg{weightflow} redistributes their weight among the
resolved cases within cells $c$,
\begin{equation}
w_i \;\leftarrow\; w_i \cdot \frac{\sum_{j\in c} w_j}{\sum_{j\in c,\ \text{resolved}} w_j},
\qquad i \in c,\ \text{resolved},
\end{equation}
so the cell total is conserved. When unknowns arrive without a roster (one
record per address) the redistribution is done at the cluster (household) level.

\paragraph{Ineligible units.}
Out-of-scope units receive weight zero. They are dropped \emph{after} the
unknown-eligibility step, so that they first absorb their share of the
unresolved weight.

\paragraph{Within-household selection.}
When one eligible person is selected per household, that person represents all
eligible persons in the household, so the weight is multiplied by the inverse of
the within-household selection probability,
$w_i \leftarrow w_i/p_i$ (or by the number of eligibles under equal-probability
selection). This restores the inclusion probabilities implied by the design.

\paragraph{Nonresponse.}
Not all sampled units respond. Under the assumption that response is ignorable
given observed auxiliaries \cite{littlerubin2002}, the responding units are
inflated to represent the nonrespondents. The adjustment can be applied at the
person level or, when a whole household is lost (for instance with no roster), at
the household level, and \pkg{weightflow} offers both weighting classes and
response-propensity models for it (Section~\ref{sec:nr}). Because the steps
compose freely, a nonresponse adjustment may appear more than once in a recipe,
for instance at the household level before within-household selection and at the
person level afterwards; within-household selection is itself a particular case
of the same inverse-probability reweighting.

\paragraph{Calibration.}
Finally the weights are aligned with known population totals of auxiliary
variables (Section~\ref{sec:calib}).

\paragraph{Finalising.}
Optional closing steps handle extreme weights and the final scale, and can be
inserted anywhere in the recipe. \textbf{Trimming} caps extreme weights by a
manual bound (a maximum or minimum weight ratio), by the Tukey far-out fence, or
by \textbf{Potter's MSE-optimal} cutoff, which chooses the threshold that
minimises an estimate of bias$^2$ plus variance \cite{potter1990} rather than a
hand-picked value. The weights can also be \textbf{normalised} (rescaled to sum
to the sample size $n$, so they average one, or to any chosen total) and rounded
to integers. Rounding that preserves the total (the largest-remainder
\code{"preserve\_total"} method of \code{step\_round}) keeps the rounded weights
summing exactly to the population size $N$, or to the calibrated control total,
which production systems often require.

\section{A declarative API}\label{sec:api}

\pkg{weightflow} separates two moments. A recipe is first \emph{defined} as an
inert object (\code{weighting\_spec()} followed by a chain of \code{step\_*()}
adjustments) and then \emph{applied} with \code{prep()}, which executes each
step in order. \code{collect\_weights()} extracts the final weights. This
\pkg{tidymodels}-style separation makes the pipeline reproducible and auditable
(it reads as a procedure) and, crucially, lets the variance estimators
re-execute the entire recipe on each replicate (Section~\ref{sec:variance}).
Every step stores per-stage diagnostics, so \code{summary()} and
\code{report\_weighting()} expose how the effective sample and the weight
distribution evolve. The package imports only base R (\pkg{stats},
\pkg{utils}, \pkg{graphics}); optional engines and the survey bridges are used
conditionally.

\section{Calibration}\label{sec:calib}

\begin{figure}[t]
\centering
\includegraphics[width=0.62\textwidth]{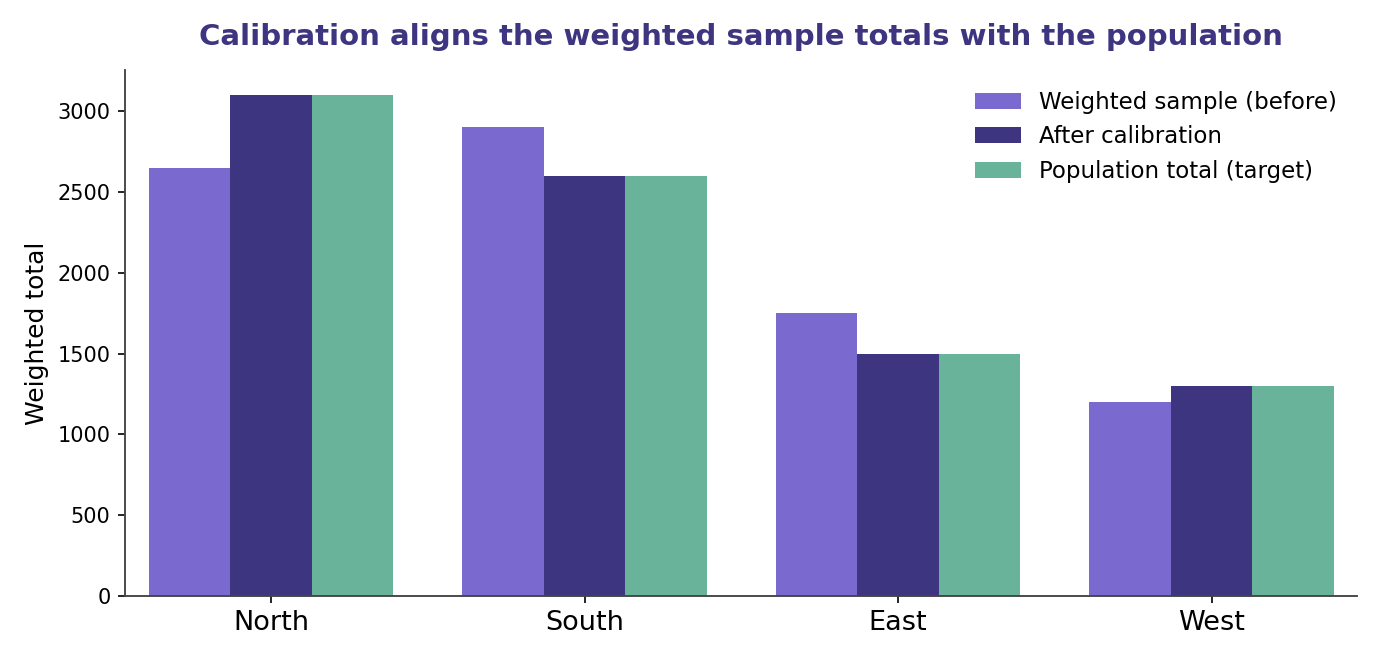}
\caption{Calibration adjusts the weights until the weighted sample totals match
the known population benchmarks.}
\label{fig:calib}
\end{figure}

Calibration finds weights $w_i$ close to the input weights $d_i$ that reproduce
known totals of auxiliaries $\mathbf{x}_i$,
\begin{equation}
\sum_{i\in s} w_i\,\mathbf{x}_i \;=\; \mathbf{X}=\sum_{i\in U}\mathbf{x}_i,
\label{eq:calib}
\end{equation}
which reduces coverage bias and, when the auxiliaries are associated with the
study variables, improves precision (Figure~\ref{fig:calib}). From a design
perspective it is the generalized regression (GREG) estimator
\cite{deville1992,deville1993}.

\pkg{weightflow} provides three methods through a single \code{step\_calibrate()}
step: \textbf{raking} (iterative proportional fitting to several margins),
\textbf{post-stratification} (to the joint cells of a cross-classification) and
\textbf{linear/GREG} (categorical and continuous auxiliaries through a model
formula). The calibrated weight is $w_i=d_i\,g_i$, with the factor $g_i$
minimising a distance to $d_i$ subject to \eqref{eq:calib}. The linear distance
gives $g_i=1+\mathbf{x}_i'\boldsymbol\lambda$; the exponential (``raking'')
distance $g_i=\exp(\mathbf{x}_i'\boldsymbol\lambda)$ keeps the weights positive
without bounds \cite{deville1993}. Bounds $c(L,U)$ truncate $g_i$ (Deville--S\"arndal). A \textbf{ridge} penalty
relaxes the exact constraint \eqref{eq:calib} to control extreme weights when
there are many, possibly collinear, auxiliaries: instead of forcing
$\sum_i w_i\mathbf{x}_i=\mathbf{X}$ exactly, it trades a small, controlled
deviation on the totals for much steadier weights,
\begin{equation}
\min_{w}\ \sum_{i} \frac{(w_i-d_i)^2}{d_i\,q_i}
\;+\; \sum_j \frac{(\hat X_j - X_j)^2}{c_j},
\qquad \hat X_j=\sum_i w_i x_{ij},
\end{equation}
where $d_i$ is the weight entering calibration (the input weight, kept distinct
from the calibrated output $w_i$), $q_i$ is a known positive tuning constant of
the distance (unit-specific, equal to one by default), and $c_j$ is the cost of
constraint $j$. In the linear case the calibration
system $A\boldsymbol\lambda=\mathbf{X}-\hat{\mathbf{X}}$ gains a penalty on its
diagonal, $\big(A+\operatorname{diag}(s/c_j)\big)\boldsymbol\lambda
=\mathbf{X}-\hat{\mathbf{X}}$, with $s$ a scale factor that makes the penalty
unit-free. A single scale-free \code{penalty} governs the overall trade-off, and
the per-constraint costs $c_j$ let the user \textbf{flexibilise} the target:
some margins can be held (almost) exactly while others are allowed to move, which
is valuable when a benchmark is uncertain or when constraints conflict.
Two further variants matter for production. \textbf{Domain-partitioned}
calibration (\code{by=}) calibrates independently within each domain, each to its
own totals, useful when a domain carries its own quantitative control total,
which is awkward to express as interactions by hand. \textbf{Integrative}
(Lemaitre--Dufour) calibration \cite{lemaitre1987}, switched on with
\code{equal\_within\_cluster = TRUE} together with a \code{cluster} identifier,
gives a single common weight to all members of a household, so that person and
household estimates stay coherent.

Population totals may be supplied in the classic model-matrix form (an intercept
plus treatment-contrast columns) or, more conveniently, as \textbf{tidy data
frames}: one row per category with a counts column, and several category columns
crossed automatically. The tidy form frees the user from dropping a reference
level and constructing the intercept by hand, which is the part of specifying
calibration totals that is most error-prone in practice. \pkg{weightflow} also
reports whether the targets were met: linear calibration warns when the
constraints are not fully satisfied (collinear or ill-conditioned auxiliaries),
raking warns on inconsistent margins or non-convergence, and post-stratification
checks the cells explicitly. A cell present in the totals but with no sampled
unit cannot be filled and raises a warning, while a cell present in the sample
but missing from the totals raises an error.

\section{Model-assisted calibration}\label{sec:modelcal}

Model calibration \cite{wusitter2001} fits a working model for each study
variable $y$, predicts $\hat y_i$ over the population, and calibrates so that the
weighted sample total of the predictions matches the population total,
\begin{equation}
\sum_{i\in s} w_i\,\hat y_i \;=\; \sum_{i\in U}\hat y_i,
\end{equation}
alongside the benchmark totals $\mathbf{X}=\sum_{i\in U}\mathbf{x}_i$ of the
auxiliaries. The result is a \textbf{hybrid} calibration: the weights
simultaneously satisfy the model-prediction constraints (which borrow efficiency
from $\hat y$) and the auxiliary benchmark totals (consistency with published
margins). When the model is linear the second constraint is implied by the first
and the estimator reduces to GREG; for nonlinear learners the prediction
constraint adds efficiency \cite{wu2003}. \pkg{weightflow} fits the working
models with generalized linear models, trees, random forests or gradient
boosting, optionally with cross-fitting; the benchmark totals $\mathbf{X}$ may
come from \code{population} or from an external source through \code{x\_totals},
and the integrative one-weight-per-household option is available here too.

\section{Nonresponse modelling}\label{sec:nr}

Under the assumption that response is ignorable given observed auxiliaries
\cite{littlerubin2002}, \pkg{weightflow} inflates the responding units to
represent the nonrespondents. It offers two families of adjustment for this step,
and both can act at the person or the household level.

\paragraph{Weighting classes.}
The sample is partitioned into cells $c$ defined by auxiliary variables, and
within each cell the responding weight is inflated to the cell total,
\begin{equation}
f_c=\frac{\sum_{i\in c} w_i}{\sum_{i\in c} r_i\, w_i},\qquad
r_i=1\ \text{if}\ i\ \text{responds},
\end{equation}
so that $w_i^{\text{nr}}=f_c\,w_i$ for a responding $i\in c$. This is the special
case of a propensity model in which the propensity is the design-weighted
response rate of the cell.

\paragraph{Response-propensity models.}
More generally the response propensity
$\hat\phi_i=\widehat{P}(\text{respond}\mid\mathbf{x}_i)$ is estimated from the
auxiliaries and the weight is adjusted at the unit level,
$w_i\leftarrow w_i/\hat\phi_i$. A single \code{engine} argument selects the
learner: logistic regression, a classification tree (CART), a random forest, or
gradient boosting (\pkg{xgboost}). Flexible learners that predict the units they
were trained on overfit the propensity and inflate the weights; optional
$k$-fold \textbf{cross-fitting} estimates each unit from a model trained on the
other folds, with the folds formed by cluster so that members of the same
household never split across folds and there is no leakage.

\paragraph{Classes from fitted propensities.}
Rather than applying $1/\hat\phi_i$ unit by unit, the estimated propensities
can be grouped into \textbf{classes}, quantile-based post-strata of the fitted
$\hat\phi_i$, with the weighting-class adjustment then applied within each
class (controlled by \code{num\_classes}). This stabilises the adjustment when
the model is imperfect and unifies the two families: weighting classes are the
propensity classes of a purely categorical model.

\section{Recipe-aware variance}\label{sec:variance}

A linearization that treats the final weights as fixed ignores that several
stages of the cascade were themselves estimated from the sample: the
unknown-eligibility redistribution, the nonresponse model, the calibration, and
any data-driven trimming. Ignoring this understates the standard error.
\pkg{weightflow} instead \textbf{re-runs the whole recipe on each replicate}, so
the replicate weights carry the variability of \emph{every} estimated stage, not
only that of the final calibration. The rescaling bootstrap \cite{raowu1988} resamples $m_h$ of the
$n_h$ primary sampling units (PSUs) within each stratum $h$ with replacement
(default $m_h=n_h-1$) and rescales unit $i$ in PSU $k$ by
\begin{equation}
\lambda_{hi}=1-\sqrt{\tfrac{m_h}{n_h-1}}+\sqrt{\tfrac{m_h}{n_h-1}}\;\frac{n_h}{m_h}\,t_{hi}^{*},
\end{equation}
where $t_{hi}^{*}$ counts how often the PSU is drawn; the factor has expectation
one and is never negative, so the recipe can be re-prepped on each replicate.
Re-preparing it on each of $B$ replicates gives replicate estimates
$\hat\theta^{*}_1,\dots,\hat\theta^{*}_B$, and the bootstrap variance of an
estimator $\hat\theta$ is their Monte Carlo variance,
\begin{equation}
\widehat V_{\mathrm{boot}}(\hat\theta)=\frac{1}{B}\sum_{b=1}^{B}
\big(\hat\theta^{*}_b-\bar\theta^{*}\big)^2,\qquad
\bar\theta^{*}=\frac{1}{B}\sum_{b=1}^{B}\hat\theta^{*}_b .
\end{equation}
The delete-a-PSU jackknife removes one PSU at a time and re-runs the recipe,
giving the stratified (JKn) estimator
\begin{equation}
\widehat V_{JK}=\sum_h \frac{n_h-1}{n_h}\sum_{i\in h}\big(\hat\theta_{(hi)}-\hat\theta_h\big)^2,
\end{equation}
or the unstratified JK1 with a single stratum. For strata with a single PSU,
which contribute no resampling variance, the units are treated as
self-representing and a warning is issued. The point weights and the replicate
weights export to \pkg{survey}/\pkg{srvyr} (as an ultimate-cluster linearization
design or a replicate-weights design), so any estimand or domain can be estimated
downstream with the recipe's uncertainty built in.

\section{Diagnostics}\label{sec:diag}

\begin{figure}[t]
\centering
\includegraphics[width=0.66\textwidth]{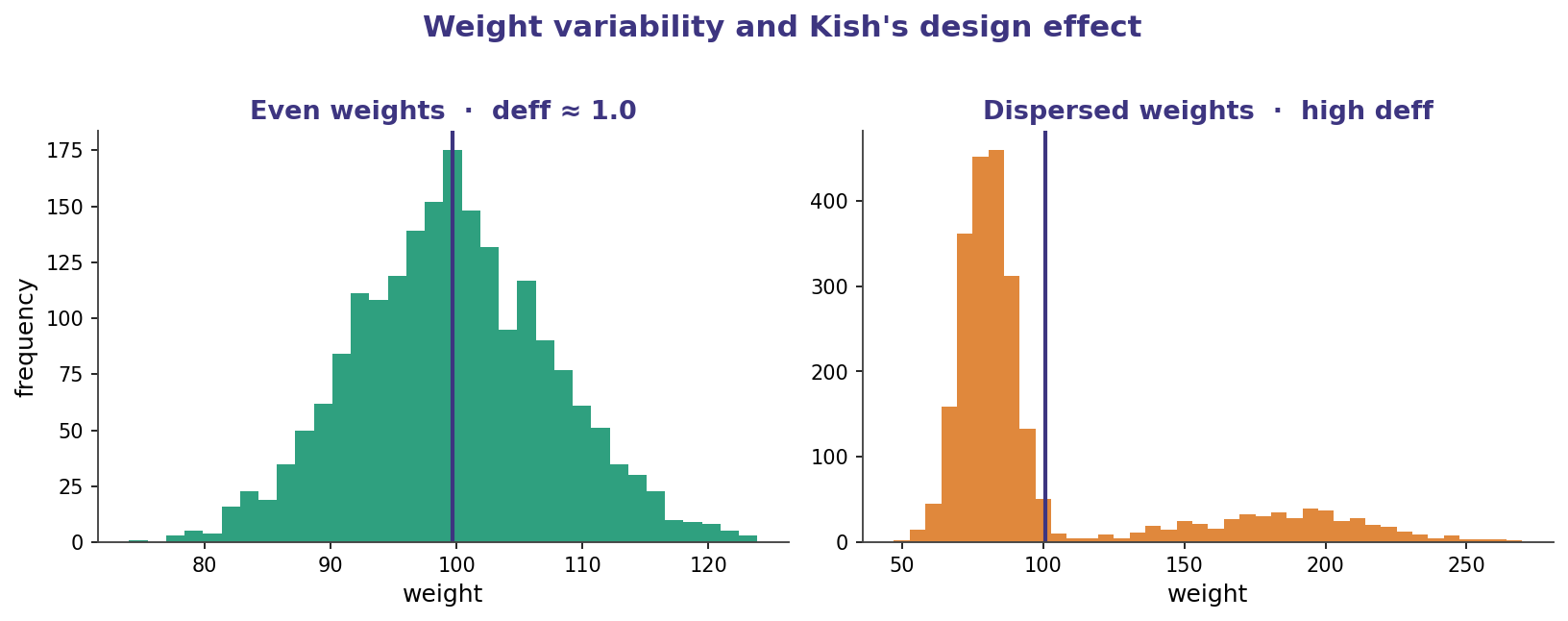}
\caption{The more dispersed the weights, the larger Kish's design effect and the
smaller the effective sample size.}
\label{fig:deff}
\end{figure}

Two questions guide diagnosis. Kish's design effect measures the precision lost
to unequal weighting,
\begin{equation}
\mathrm{deff}=1+CV^2(w)=\frac{n\sum_i w_i^2}{\big(\sum_i w_i\big)^2},\qquad
n_{\mathrm{eff}}=\frac{n}{\mathrm{deff}},
\end{equation}
so more dispersed weights buy a smaller effective sample
(Figure~\ref{fig:deff}). The representativity of the response is summarised by the
R-indicator \cite{schouten2009},
\begin{equation}
R=1-2\,S,\qquad
S=\sqrt{\tfrac{1}{N-1}\sum_i w_i\,(\hat\phi_i-\bar\phi)^2},\qquad
\bar\phi=\frac{1}{N}\sum_i w_i\,\hat\phi_i,
\end{equation}
where $\hat\phi_i$ is the estimated response propensity of unit $i$
(Section~\ref{sec:nr}), $w_i$ its design weight, $\bar\phi$ the weighted mean
propensity, and $S$ the (design-)weighted standard deviation of the propensities.
Values of $R$ near one indicate a more representative response and less
nonresponse-bias risk. When the recipe
includes a nonresponse step, \code{summary()} and \code{report\_weighting()}
report $R$ and its unconditional partial indicators automatically, pointing to
the variable that drives the lack of representativity.

\section{Illustration: a full cascade}\label{sec:example}

The bundled \code{sample\_one} data set is a multistage select-one design. A
complete cascade (unknown eligibility and household nonresponse handled at the
household level, within-household selection, person nonresponse in demographic
cells, and raking to region and sex totals) reads as one pipe:

\begin{quote}\small
\begin{verbatim}
rec <- weighting_spec(dat, base_weights = pw) |>
  step_unknown_eligibility(unknown = unknown_elig, by = "region",
                           cluster = "household_id") |>
  step_drop_ineligible(ineligible = ineligible) |>
  step_nonresponse(respondent = hh_responded, method = "weighting_class",
                   by = "region", cluster = "household_id") |>
  step_select_within(prob = p_within) |>
  step_nonresponse(respondent = responded, method = "weighting_class",
                   by = c("region", "sex", "age_grp")) |>
  step_calibrate(method = "raking",
                 margins = list(region = c(table(population$region)),
                                sex    = c(table(population$sex)))) |>
  prep()

summary(rec)                       # per-stage diagnostics + R-indicator
w    <- collect_weights(rec)$.weight
boot <- bootstrap_weights(rec, replicates = 200, strata = "region",
                          psu = "psu", seed = 1)
boot_mean(boot, "income")          # estimate, SE and CI

# or hand the replicate weights to survey / srvyr and analyse as usual:
rep_design <- as_svrepdesign(boot)            # a svyrep.design object
svymean(~income, rep_design)                  # design + cascade variance
\end{verbatim}
\end{quote}

\code{summary()} prints, for each stage, the active sample size, sum of weights,
coefficient of variation, Kish design effect and effective sample size, and, as
the recipe adjusts for nonresponse, the R-indicator and its partials. The
recipe-aware bootstrap then yields standard errors that carry the variability of
every adjustment. Because those replicate weights already encode both the
sampling design (resampled PSUs within strata) and the weighting cascade
(re-estimated on each replicate), \code{as\_svrepdesign()} hands them to
\pkg{survey}/\pkg{srvyr} so that any estimator, domain or model can be fitted
downstream with the recipe's uncertainty carried automatically, exactly as one
would analyse ordinary replicate weights.

\section{A real-data application: the Uruguayan ECH}\label{sec:ech}

To exercise the full pipeline on production-scale data, we apply it to the 2019
Uruguayan continuous household survey (Encuesta Continua de Hogares, ECH),
released as open microdata by INE: about 79{,}000 person records expanding to a
population above three million. The public file contains only eligible
respondents, so we induce ineligibility, unknown eligibility and nonresponse at
the dwelling level in a seeded, reproducible way, with a missing-at-random
mechanism in which poorer strata and younger households respond less. Because the
base weight is set to the published ECH person weight, the base-weighted poverty
rate reproduces the population value exactly, giving a \emph{known truth}
$p=0.0876$ against which every stage can be checked.

The recipe redistributes the unknown-eligibility weight, drops the ineligibles,
adjusts for nonresponse by weighting classes within strata (all at the household
level through \code{cluster}), calibrates to age, sex and region with a single
weight per household (\code{equal\_within\_cluster = TRUE}), and closes with
Potter trimming and total-preserving rounding.

\begin{figure}[t]
\centering
\includegraphics[width=0.78\textwidth]{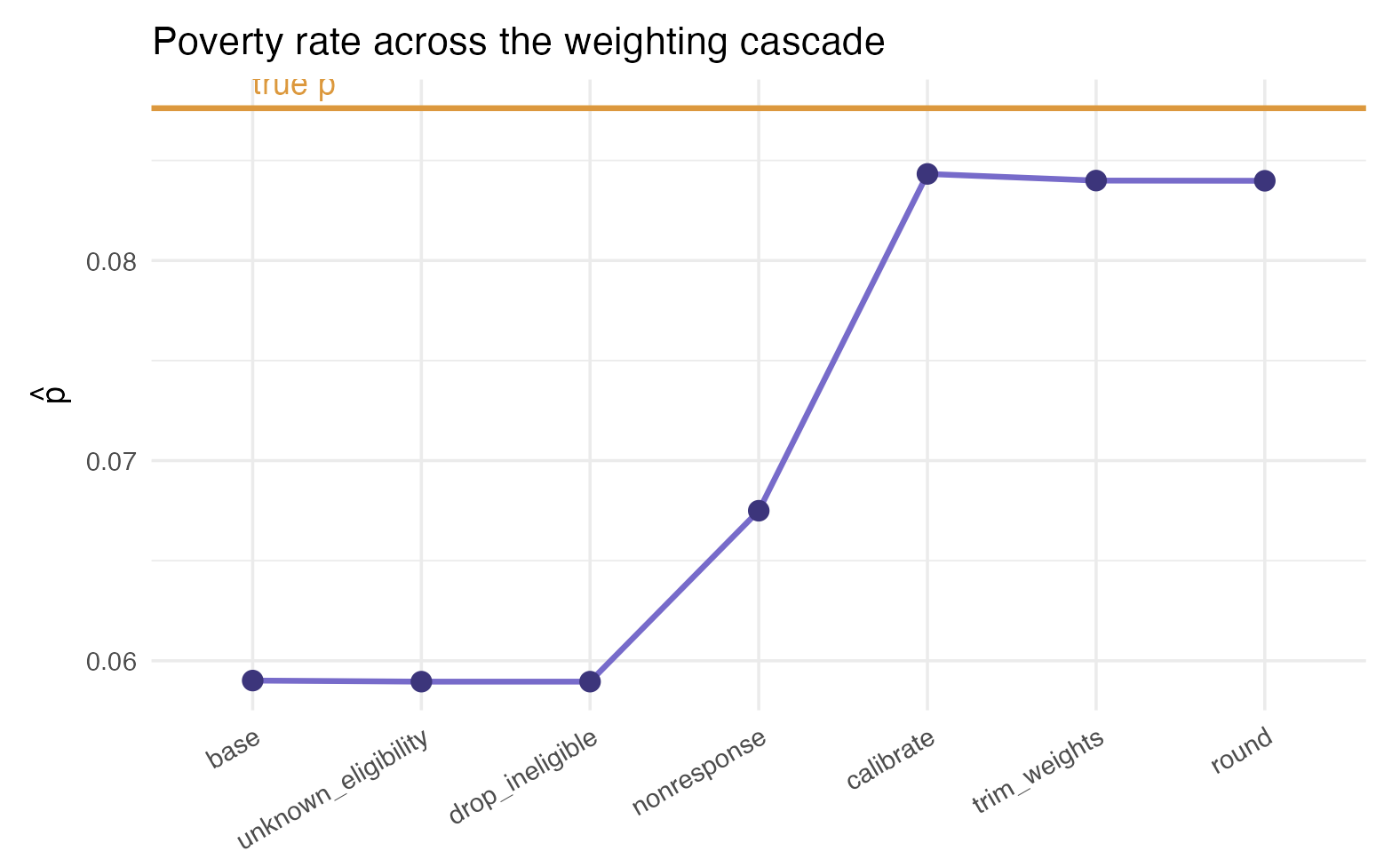}
\caption{Poverty-rate estimate at each stage of the ECH cascade. Each adjustment
moves the estimate toward the known truth (amber line).}
\label{fig:ech-cascade}
\end{figure}

Figure~\ref{fig:ech-cascade} tracks the estimated poverty rate along the cascade.
The naive base estimate is badly biased downward ($\hat p=0.059$ against
$p=0.0876$) because the poor responded less; the nonresponse adjustment recovers
the between-strata part ($\hat p=0.068$), and calibration to the age, sex and
region margins recovers most of the rest ($\hat p=0.084$), landing close to the
truth. A small residual gap is honest and expected, since the adjustments use
only the frame and the calibration margins. The Kish design effect rises from
$1.08$ at base to $1.42$ after calibration, the price of the unequal weighting.

\begin{figure}[t]
\centering
\includegraphics[width=0.78\textwidth]{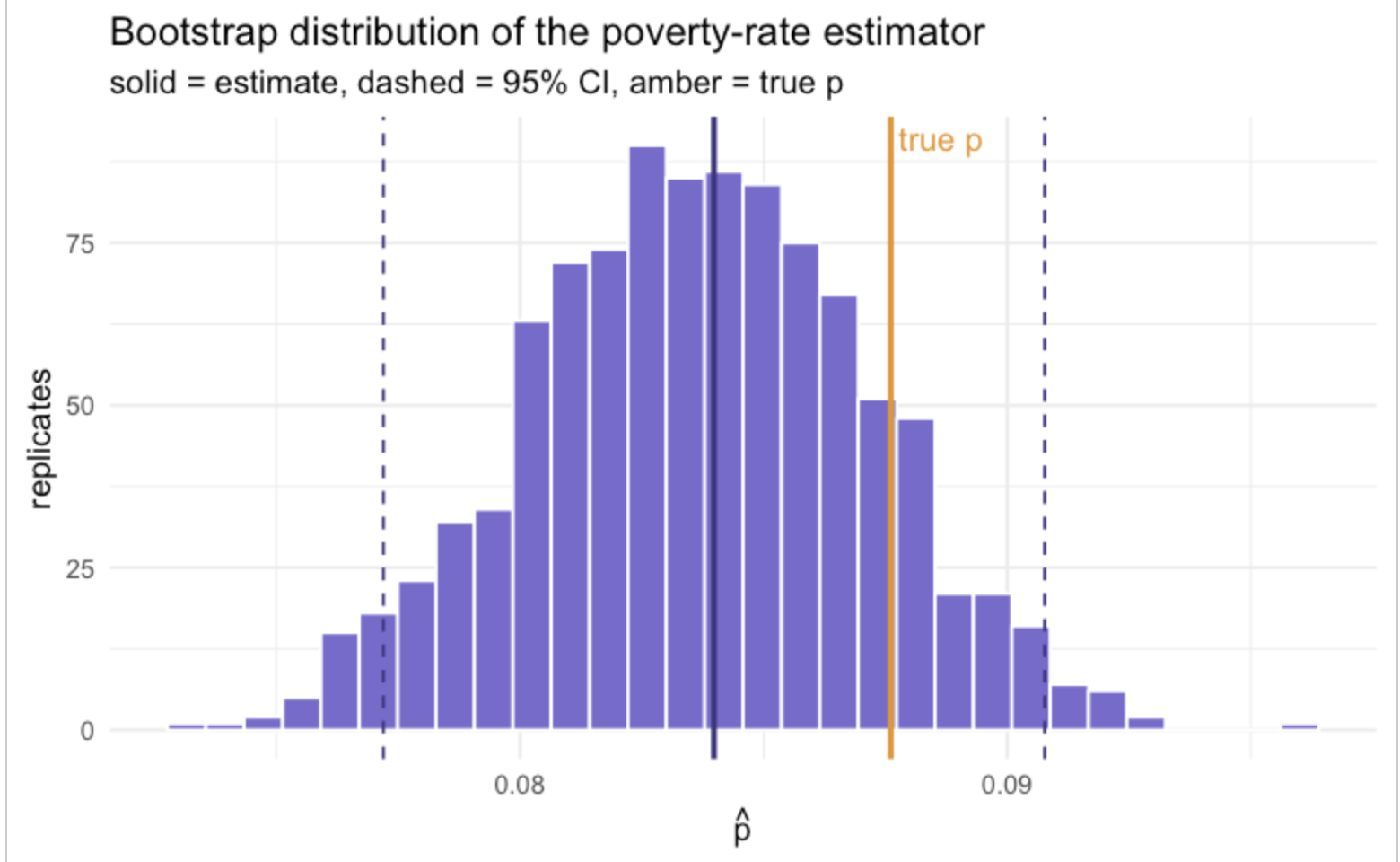}
\caption{Bootstrap distribution of the poverty-rate estimator over $1{,}000$
recipe-aware replicates. Solid line: point estimate; dashed: $95\%$ interval;
amber: the true rate, which the interval covers.}
\label{fig:ech-boot}
\end{figure}

Design-based inference uses the recipe-aware bootstrap: $1{,}000$ replicates that
resample PSUs within strata and re-apply the entire recipe. The resulting $95\%$
interval covers the true poverty rate (Figure~\ref{fig:ech-boot}), so the
pipeline recovers the parameter with honest uncertainty.

The run also serves as a light benchmark. On a laptop (Apple M4), preparing the
full recipe on about 79{,}000 records takes $0.45$ seconds, so the point estimate
and all its diagnostics are essentially instant. The bootstrap is the heavy part,
because it re-applies the whole recipe on each replicate: $1{,}000$ replicates
take about $344$ seconds (under six minutes), which puts full design-based
inference on real survey data within routine reach on a single machine.

\section{Validation and related software}\label{sec:validation}

On the methods they share, \pkg{weightflow} reproduces \pkg{survey}: on the same
starting weights and control totals, post-stratification, raking and linear
calibration return the same weights (to numerical tolerance), and the
delete-a-PSU jackknife matches \pkg{survey}'s replicate jackknife for totals.
These agreements are checked in the package's test suite and in a dedicated
validation vignette. Table~\ref{tab:compare} situates \pkg{weightflow} among
related CRAN packages: its distinctive combination is the \emph{full cascade as
one recipe} together with \emph{recipe-aware replicate variance}, implemented
without hard dependencies.

\begin{table}[t]
\centering
\small
\caption{\pkg{weightflow} relative to related R packages (indicative, not a
ranking; packages differ in focus).}
\label{tab:compare}
\begin{tabular}{lccccc}
\toprule
Feature & \pkg{weightflow} & \pkg{survey} & \pkg{sampling} & \pkg{surveysd} & \pkg{svrep} \\
\midrule
Full cascade as one recipe        & \checkmark & -- & -- & -- & -- \\
Raking / post-strat. / GREG       & \checkmark & \checkmark & partial & \checkmark & -- \\
ML nonresponse + cross-fitting    & \checkmark & -- & -- & -- & -- \\
Hybrid model-assisted calibration & \checkmark & -- & -- & -- & -- \\
Recipe re-applied on each replicate & \checkmark & -- & -- & partial & partial \\
Replicate weights to survey/srvyr & \checkmark & \checkmark & -- & \checkmark & \checkmark \\
R-indicators (automatic)          & \checkmark & -- & -- & -- & -- \\
No hard dependencies (base R)     & \checkmark & -- & -- & -- & -- \\
\bottomrule
\end{tabular}
\end{table}

\section{Discussion}\label{sec:discussion}

\pkg{weightflow} builds on established adjustments rather than proposing new
point estimators, but it contributes a capability that, to our knowledge, no
existing package offers: a replicate variance that re-executes the \emph{entire}
recipe on every replicate, so the uncertainty of every estimated stage (unknown
eligibility, ineligible dropping, within-household selection, nonresponse and
calibration) reaches the standard errors, not only that of the final
calibration. Together with the auditable, dependency-free recipe abstraction and
the ML-based, hybrid model-assisted calibration, this makes \pkg{weightflow} more
than a repackaging of known methods. It suits national statistical offices, where
auditability, reproducibility and longevity matter as much as the methods
themselves. The recipe abstraction is extensible:
new adjustments are added as \code{step\_*()} constructors and inherit the
variance machinery for free. Ongoing work includes richer handling of
single-PSU strata (beyond the self-representing default) and performance at
production scale. The package is on CRAN, developed openly, and covered by an
extensive test suite (invariants, guardrails, snapshot and cross-platform
continuous integration).

\section*{Availability}
\pkg{weightflow} is available on CRAN
(\url{https://CRAN.R-project.org/package=weightflow}); source, documentation and
articles are at \url{https://jpferreira33.github.io/weightflow/}. It is released
under the MIT licence.

\end{document}